# İyon akım rektifikasyonuna bağlı pH duyarlı nanoprob üretimi ve karakterizasyonu

# Fabrication and characterization of pH responsive nanoprobes based on ion current rectification

Mustafa ŞEN

Biyomedikal Mühendisliği, İzmir Kâtip Çelebi Üniversitesi, İzmir, TURKEY
mustafa.sen@ikc.edu.tr

*Özetçe*—Bu çalışma kapsamında bovin serum albümin (BSA) ile modifiye edilmiş cam nanopipetlerin gösterdiği iyon akım rektifikasyonu farklı pH değerlerine sahip solüsyonlarda araştırılmıştır. Temel olarak iyon akım rektifikasyonu nanopipetlerde asimetrik *I-V* eğrisi olarak gözlemlenmektedir. İyon akım rektifikasyonu, nanopipetlere uygulanan potansiyelin yönüne bağlı ölçülen akımın farklılık göstermesi durumudur. Modifikasyonu yapılan nanopipetlerle elde edilen sonuçlar, nanopipetlerin modifikasyon sonrasında pH bağımlı iyon akım rektifikasyon davranışına sahip olduğunu göstermektedir. Önerilen üretim stratejisiyle hücre içi pH ölçümü yapabilecek pH duyarlı nanoprobların kolayca üretilebileceği düşünülmektedir.

*Anahtar Kelimeler — Nanoprob, pH, iyon akım rektifikasyonu, yapay membran*

*Abstract*—In this study, we investigated the ionic current rectification of glass nanopipettes modified with bovine serum albumin – glutaraldehyde (BSA-GA) artificial membrane using solutions with various pHs. Ionic current rectification is a phenomenon that is observed with nanopores as asymmetric I-V curves, where the ionic currents recorded through a nanopore differ at the same magnitude of applied electrical potentials biased with opposite polarities. The results clearly showed that modifying the tip of a nanopipette results in a pH dependent ionic current behavior. The proposed strategy is a facile method for fabrication of a pH responsive nanoprobe that has a potential for intracellular pH measurement.

## I. GİRİŞ

Hücre membranlarında bulunan nanoporlarda gerçekleşen taşınım çok sayıda biyolojik işlem için büyük bir öneme sahiptir. Bu nanoporların kullanımı DNA sekanslama gibi çok sayıda biyoteknolojik uygulamaya kapı aralamıştır [1]. Biyolojik nanoporlardan esinlenerek katı hal nanoporlar ile geliştirilen biyolojik ve kimyasal analizler günümüzde güçlü ve gelecek vadeden bir teknoloji haline gelmiştir. Temel olarak belli bir potansiyel farkı uygulanan nanoporlardan geçen iyonik akım nanoporun içine ve dışına ayrı ayrı yerleştirilen elektrotlar yardımıyla ölçülmektedir. İyonik akımda gözlemlenen farklılık nanopordan geçen iyonlarla veya modifiye edilmiş nanoporlarda bulunan tanıma bölgelerine bağlanan moleküller ile ilişkilidir [2-4]. Genellikle, nispeten büyük porlara sahip katı hal nanoporları pH, antijen veya inhibitör gibi biyolojik uyarıcılara karşı duyarlı veya seçici değillerdir. Dolayısıyla bu tip nanoproblar deteksiyonu arzulanan moleküllere özgü şekilde modifiye edilmeleri gerekmektedir. Sinyal yükseltme ve her hangi bir biyobelirteç kullanım gereksiniminin olmaması nano porlarla algılamayı ilgi çekici hale getirmektedir. Katı hal nanoporları genellikle iki şekilde üretilmektedir. Bunlardan ilki ve en çok kullanılanı litografi tekniği ile dağlamadır [5-7]. Bu teknik için pahalı makinelere ve özel laboratuvarlara gereksinim olması bu tekniğin yaygın ve ucuz kullanımını sınırlamaktadır. Kullanılan diğer bir teknik cam kılcal tüplerinin mikro çekme makinesi ile ısı altında çekilmesidir [8-11]. Bu teknik nanolitografiye nazaran daha ucuz ve daha az iş yükü gerektirmektedir. Ayrıca, mikroçekme parametreleri değiştirilerek nanoporların

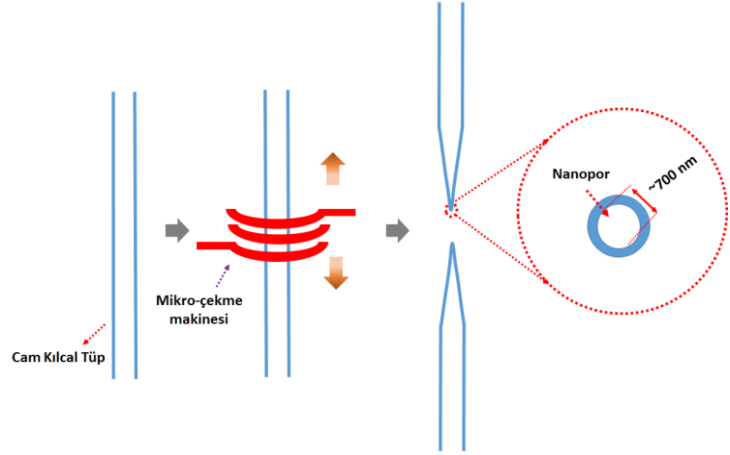

**Şekil 1.** Nanopipetlerin mikroçekme makinesiyle üretimi boyutları ve şekilleri yüksek çözünürlükte manipüle edilebilmektedir.

İyon akım rektifikasyonu kısaca nanoporlara uygulanan potansiyelin yönüne bağlı olarak elde edilen akımın farklılık göstermesi hali olarak tanımlanabilir. Bu çalışma kapsamında bovin serum albümin (BSA) ile modifiye edilmiş cam nanopipetlerin gösterdiği iyon akım rektifikasyonu farklı pH değerlerine sahip solüsyonlarda araştırılmıştır. Ayrıca, modifiye edilmemiş nanoprobların aynı koşullar altında gösterdikleri iyon akım rektifikasyon davranışları modifiye edilmiş nanoprobların ki ile karşılaştırılarak BSA-glutaraldehit yapay membranın iyon akım rektifikasyonuna etkisi ortaya konmuştur.

## II. YÖNTEM VE SONUÇLAR

Bu çalışmada nanopor içeren nanopipetler ucuz ve daha kolay olması nedeniyle mikroçekme makinesi ile üretilmişlerdir. "Patch clamp" cam kılcal tüpleri (PG10165-4, World Precision Instrument, ABD) mikro çekme makinesinde (PC-10, Narishige, Japonya) var olan 2 aşamalı çekme modu kullanılarak iğne uç tipli nanopipetler üretilmiştir (1. çekme ısısı: ˚60 C; 2. çekme ısısı: 39 ˚C) (Şekil 1). Üretilen nanopipetlerin SEM resimlerinde yapılan ölçümlerden yaklaşık 350 nm seviyesinde yarıçapa sahip oldukları saptanmıştır. Çalışma boyunca aynı boyutlara sahip nanoprob üretimi için aynı üretim parametreleri kullanılmıştır. Daha sonrasında nanopipet ucu BSA-glutaraldehit yapay membranı ile modifiye edilmiştir. BSA-glutaraldehit yapay membranı kontrollü ilaç iletimi ve biyosensörlerin biyotanıma ünitesinde enzim immobilizasyonu gibi çeşitli uygulamalarda yaygın olarak kullanılmaktadır [12-14]. Temel olarak, glutaraldehitte bulunan oldukça reaktif aldehit grupları BSA proteininde bulunan serbest amin grupları ile çapraz bağlanarak dakikalar içerisinde yapay membranın oluşumunu sağlamaktadır. Nanopipetlerin uç kısmının BSA-glutaraldehit yapay membranı ile modifiye edilmesinin akabinde bu probların pH duyarlılığı farklı pH değerlerine sahip solüsyonlarda test edilmiştir. Bu noktada nanopipetlerin içi PBS (pH 7) ile doldurulmuş ve sonrasında bu nanopipetlerin farklı pH değerlerine sahip PBS solüsyonlarında iyon rektifikasyon davranışları araştırılmıştır. İyon akım ölçümü için 2 adet Ag/AgCl elektrot kullanılmıştır; bunlardan ilki nanopipet içine diğeri ise nanoprobun içine daldırıldığı farklı pH değerlerine sahip PBS solüsyonlarının içine yerleştirilmiştir. Bu iki elektrot arasında ve aynı zamanda nanopipetin ucundaki naopordan geçen iyon akımı bir potentiyostat yardımıyla ölçülmüştür (Autolab PGSTAT101, Metrohm, İsviçre). İyon akım ölçümünde nanoproblara uygulanan potansiyel +0.5 ile -0.5 V arasında taranarak tipik $I$-$V$ eğrileri elde edilmiştir. BSA-glutaraldehit yapay membranı ile modifikasyonun iyon akım rektifikasyonuna etkisinin daha net olarak belirlenebilmesi için aynı koşullar altında modifiye edilmemiş nanoprobların iyon rektifikasyon davranışları da araştırılmıştır. Şekil 2'den anlaşılacağı gibi modifiye edilmemiş nanoproblar farklı pH değerlerine sahip solüsyonlarda herhangi bir iyon rektifikasyon davranışı göstermemişlerdir (Şekil 2A,B). Önceden belirtildiği

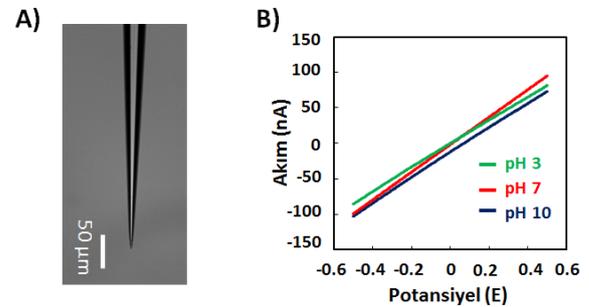

**Şekil 2.** Modifiye edilmemiş bir nanopipetin optik resmi ve farklı pH değerlerine sahip PBS çözeltilerinde gösterdiği iyon rektifikasyon davranışları

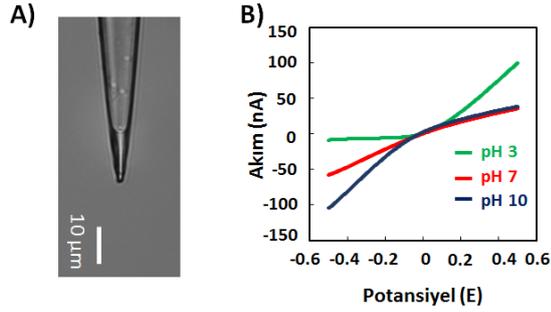

**Şekil 3.** BSA-glutaraldehit ile modifiye edilmiş bir nanopipetin optik resmi ve farklı pH değerlerine sahip PBS çözeltilerinde gösterdiği iyon rektifikasyon davranışları gibi nispeten geniş por çapına sahip nanoprobların iyon akım rektifikasyon davranışı gösterebilmeleri için uygun moleküllerle modifiye edilmeleri gerekmektedir. Buna karşın BSA-glutaraldehit yapay membranı ile modifiye edilmiş nanoprobar içine daldırılan PBS solüsyonunun pH değerine bağlı farklı iyon akım rektifikasyon davranışı göstermiştir (Şekil 3A,B). Temel olarak farklı pH değerlerinde BSA proteininin net yükü değişmekte ve bu da farklı potansiyel değerlerinde iyonların nanopordan geçişini etkilemektedir. Diğer bir değişle düşük pH değerlerinde BSA pozitif yüklü hale gelmekte ve bu da negatif potansiyel değerlerinde nanopordan geçen katyonların transportunu sınırlamaktadır.

Kısaca, bu çalışma kapsamında BSA-glutaraldehit yapay membranı ile modifiye edilmiş nanoprobların iyon akım rektifikasyon davranışları farklı pH değerlerine sahip PBS solüsyonlarında araştırılmıştır. Ayrıca, modifiye edilmemiş nanoprobların aynı koşullar altında iyon akım rektifikasyon davranışları elde edilerek modifiye edilmiş nanoprobarla karşılaştırılmıştır. Modifiye edilmemiş nanoprobların farklı pH değerlerine sahip PBS solüsyonlarında iyon akım rektifikasyon davranışı göstermemelerine karşın yapay membran ile modifiye edilmiş nanoprobar net olarak pH bağımlı iyon akım rektifikasyonu göstermişlerdir. Bildiğimiz kadarıyla bu çalışma nanopipetlerin iyon akım rektifikasyonunun ayarlanmasında yapay membranların kullanımını gösteren ilk çalışmadır. Önerilen üretim stratejisiyle hücre içi pH ölçümü yapabilecek pH duyarlı nanoprobların kolayca üretilebileceği düşünülmektedir.

## III. TEŞEKKÜR